\documentclass{article}

\usepackage[english]{babel}

\usepackage[letterpaper,top=2cm,bottom=2cm,left=3cm,right=3cm,marginparwidth=1.75cm]{geometry}

\usepackage{amsmath}
\usepackage{authblk}
\usepackage{graphicx}
\usepackage[colorlinks=true, allcolors=blue]{hyperref}
\usepackage[colorlinks=true, allcolors=blue]{hyperref}
\usepackage{adjustbox}
\usepackage{array}
\usepackage{tablefootnote}
\usepackage{comment}
\usepackage{subcaption}
\usepackage{tabularx}

\usepackage[super]{nth}

\title{Evaluating state-of-the-art cloud quantum computers for quantum neural networks in gravitational waves data analysis}
\author[1,2]{Maria-Catalina Isfan }
\author[1]{Laurentiu-Ioan Caramete}
\author[1]{Ana Caramete}
\affil[1]{Institute of Space Science – INFLPR Subsidiary, 409 Atomistilor, Magurele, Romania}
\affil[2]{Doctoral School of Physics, Faculty of Physics, University of Bucharest, 405 Atomistilor, Magurele, Romania}

\begin{document}
\maketitle

\begin{abstract}
In this work, we explore the possibility of using quantum computers provided for usage in cloud by big companies (such as IBM, IonQ, IQM Quantum Computers, etc.) to run our quantum neural network (QNN) developed for data analysis in the context of LISA Space Mission, developed with the Qiskit library in Python. Our previous work demonstrated that our QNN learns patterns in gravitational wave (GW) data much faster than a classical neural network, making it suitable for fast GW signal detection in future LISA data streams. Analyzing the fees from hardware providers like IBM Quantum, Amazon Braket and Microsoft Azure, we found that the fees for running the first segment of our QNN sum up to \$2000, \$60000, and \$1000000 respectively. Using free plans, we succeed to run the 3-qubit feature map of the QNN for one random data sample on {\fontfamily{qcr} \selectfont ibm\_kyoto} and {\fontfamily{qcr}\selectfont IQM Quantum Computers\_Garnet} quantum computers, obtaining a fidelity of 99\%; we could also run the first prediction segment of our QNN on {\fontfamily{qcr} \selectfont ibm\_kyoto}, implemented for 4 qubits, and obtained a prediction accuracy of 20\%. We queried providers such as IBM Quantum, Amazon Braket, Pasqal, and Munich Quantum Valley to obtain access to their plans, but, with the exception of Amazon Braket, our applications remain unanswered to this day.  Other major setbacks in using the quantum computers we had access to included Qiskit library version issues (as in the cases of IBM Quantum and IQM Quantum Computers) and the frequent unavailability of the devices, as was the case with the Microsoft Azure provider.

All the results presented in this paper were accumulated in 2024. 
\end{abstract}

\section{Introduction}

% Quantum computing has emerged as a promising paradigm for achieving speedups in problems that are challenging for classical computers. By leveraging superposition and entanglement, quantum algorithms can process vast amounts of data in parallel, offering potential breakthroughs in optimization, simulation, and analysis. As space missions like the Laser Interferometer Space Antenna (LISA) produce increasingly complex and data-intensive measurements, quantum computing provides an opportunity to accelerate data processing and improve the precision of astrophysical analyses.

It was demonstrated that quantum computing can speedup some computational processes, offering potential breakthroughs in optimization, simulation of quantum chemistry and data processing. This potential advantage of quantum computing algorithms, may improve data analysis pipelines for space missions such as the Laser Interferometer Space Antenna (LISA) \cite{LISAwhitepaper}, \cite{LISAredbook}. Because such experiments provide intricate and complex data, quantum computing provides an opportunity to accelerate data processing, especially through quantum machine learning (QML) \cite{QML}, \cite{QMLstateoftheart} approaches, such as quantum neural networks (QNNs) \cite{QNN}.

% In our earlier work, we were among the first to explore the use of QAI techniques, and specifically quantum neural networks (QNNs), for data analysis in the context of the LISA Space Mission. Our approach demonstrated that QNNs can successfully classify and detect gravitational wave (GW) signals embedded in realistic LISA-like noise environments, achieving high accuracy and low latency in detection tasks. These results established QAI as a viable framework for enhancing the efficiency and scalability of GW data analysis, opening the path for further exploration of quantum-enhanced methods in space-based observatories.

In our earlier work \cite{isfan2023}, \cite{isfan2025}, we investigated the use of QNNs for GW detection in the context of LISA data analysis. Our results showed that this method is viable and improves both efficiency and scalability. Moreover, it also preserves the low-latency requirements, paving the way for further studies on quantum-enhanced machine learning methods for GW data analysis.

In our earlier work \cite{isfan2023}, \cite{isfan2025}, we investigated the use of QNNs for GW detection in the context of LISA data analysis. Working with the Sangria dataset \cite{sangria}, we showed that our QNN can learn the subtle patterns in data, being able to discern between GW signals immersed in noise and noise with an accuracy exceeding 98\%. Following this result, the QNN detected 5 out of the 6 GW coalescences present in the Sangria dataset and their coalescence times. The QNN features 4 qubits, which are all measured at the end of the quantum circuit, and 64 trainable parameters.

% In this paper, we take a complementary and more hardware-oriented perspective. We explore the feasibility of deploying our QNN, built in Python using the Qiskit library, on real quantum processors offered in the cloud (by providers such as IBM, IonQ, IQM Quantum Computers, and others). Our goal is to bridge the gap between proof-of-principle QNN models and their execution on actual quantum hardware. We benchmark the fidelity of quantum circuits as they run on these devices, quantify the monetary cost of executing the full QNN workflows (including compilation, queuing, and runs), and analyze the overheads and interactions required when interfacing with the hardware provider APIs. All experiments and results presented here were collected in 2024.

In this paper, we take a complementary, hardware-oriented approach. We investigate the possibility and feasibility of deploying our QNN, implemented in Python with the Qiskit library \cite{qiskit}, on real quantum processors accessible in cloud through platforms such as those provided by IBM Quantum \cite{ibm_quantum}, Microsoft Azure \cite{azure_quantum} and Amazon Braket \cite{amazon_braket}. We summarize the available offerings, outline our interactions with the providers, and present the results from running our QNN on actual quantum hardware. Running our algorithms on real quantum hardware is essential for several reasons. First, it allows us to assess whether current state-of-the-art devices can reliably support the LISA-relevant QNN architectures. Second, hardware execution provides direct insight into practical limitations, including runtime costs, noise levels, and device-specific constraints, which are difficult to capture through simulations alone. Finally, the knowledge gained from these tests is broadly transferrable to other space-mission scenarios, offering a realistic baseline for evaluating the feasibility of quantum-enhanced data-analysis pipelines in future applications.

All experiences and results presented here were collected in 2024.

This paper is structured as follows. In Section \ref{sec:program}, we present the quantum computing programs considered for execution on quantum hardware. Section \ref{sec:approach} describes the costs of running these programs on different cloud-based quantum computers, the execution results, and our interactions with the providers. In Section \ref{sec:qiskit}, we discuss the Qiskit-related issues that hindered access to the quantum systems. Finally, the last section presents a discussion of the results and the limitations of cloud access to quantum computers.

\section{The Quantum Computing Program We Consider to Run on a Quantum Computer} \label{sec:program}

\begin{figure}
    \centering
    \includegraphics[width=\linewidth]{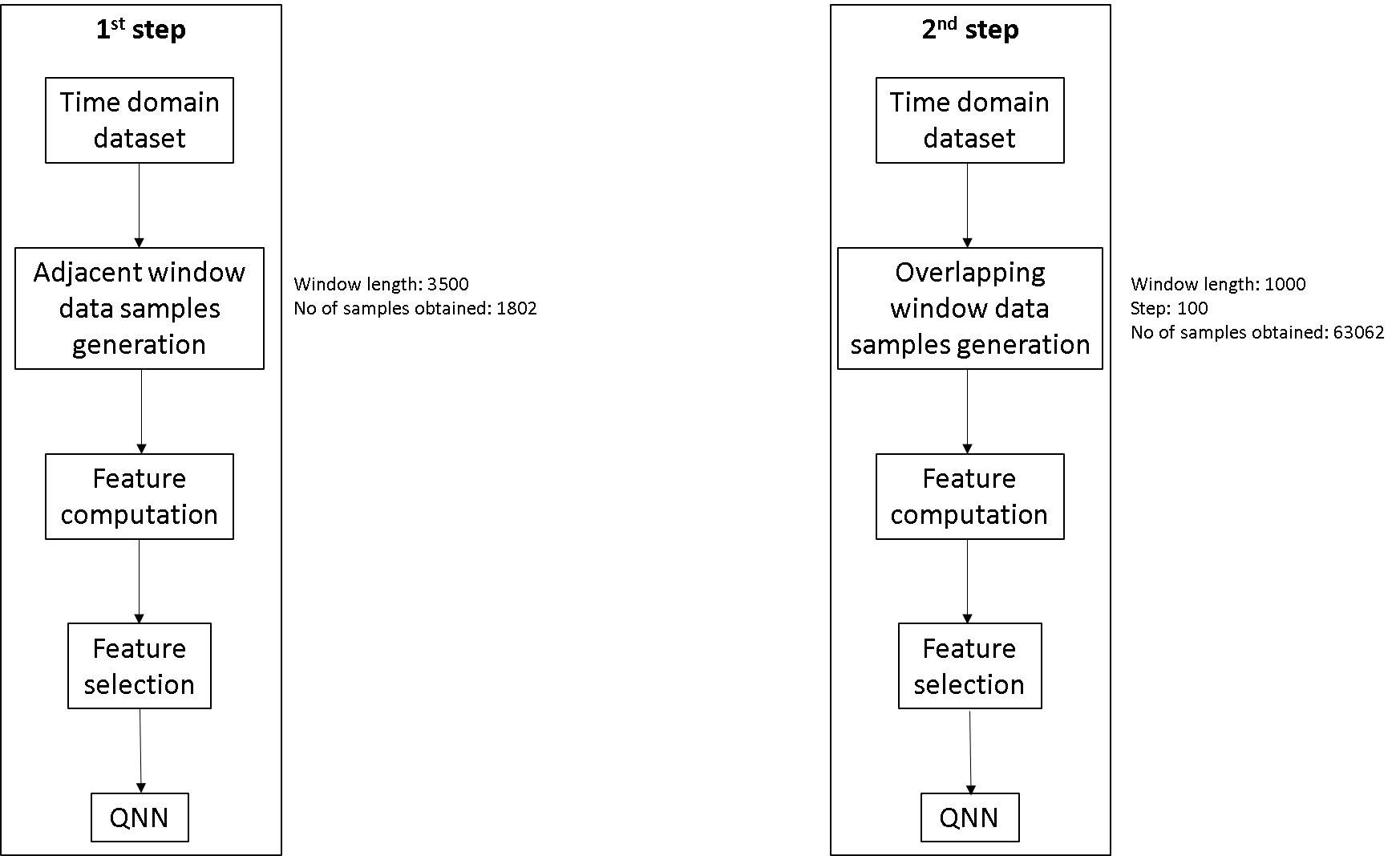}
    \caption{Data samples generation flow for each step. Data samples are generated from a time domain dataset using sliding windows: adjacent windows and overlapping windows for the \nth{1} step and for the \nth{2} step respectively.}
    \label{fig:data_samples_generation_diagram}
\end{figure}

We will present the QNN which we developed in our work in gravitational waves data analysis, in the context of LISA Space Mission. It is a Variational Quantum Classification (VQC) based QNN, involving a feature map for data encoding, an ansatz featuring trainable parameters, qubits measurement at the end of the quantum circuit and classical optimization. The quantum circuit as well as the workflow is developed with Qiskit \cite{qiskit}. Input data is encoded using the \nth{1} order Pauli expansion circuit, and the ansatz consists of a Pauli Two-Design circuit \cite{paulitwodesign1} \cite{paulitwodesign2} which is repeated 4 times (Fig.\ref{fig:IBM_sQNN}). At the end of the whole circuit, all 4 qubits are measured. This work is detailed in \cite{isfan2025}. In that paper, we state that the QNN is trained in two steps, each step employing slightly different data pre-processing methods (Figure\ref{fig:data_samples_generation_diagram}). As the first step uses a smaller set of data samples and the second step uses a bigger set of data samples, we can consider that each training step involves a short QNN (sQNN), and a long QNN (lQNN), respectively. The sQNN and the lQNN have the same architecture, the only difference being the size of the dataset they operate with. Here, we list the essential information about the sQNN and the lQNN (Tab.\ref{tab:qnns}).

% Define a centered column type with width
\newcolumntype{C}[1]{>{\centering\arraybackslash}m{#1}}
\begin{table}[htbp]
    \centering
    \begin{adjustbox}{max width=\textwidth}
    \small
        \centering
        \begin{tabular}{|C{0.4\linewidth}|C{0.3\linewidth}|C{0.3\linewidth}|}
            \cline{2-2} \cline{2-3} 
            \multicolumn{1}{c|}{} & \textbf{sQNN} & \textbf{lQNN} \\
            \hline
            \textbf{\# qubits} &  3 or 4 \tablefootnote{Very similar sQNN accuracies are obtained when the quantum circuit of the sQNN is run using 3 or 4 qubits.} &  4 \\
            \hline
            \textbf{total circuit depth} & 16 (in the case with 3 qubits) or 13 (in the case with 4 qubits) & 13 \\
            \hline
            \textbf{\# gates} & 110 (in the case with 3 qubits) or 124 (in the case with 4  qubits) & 124 \\
            \hline
            \textbf{\# epoches} & $\approx$160 & $\approx$170 \\
            \hline
            \textbf{optimizer} & Cobyla & Cobyla \\
            \hline
            \textbf{\# samples} & 5406 & 126124 \\
            \hline 
            \textbf{\# training samples} & 3604 & 63062 \\
            \hline
            \textbf{\# prediction samples} & 1082 & 63062 \\
            \hline
            \textbf{\# trainable parameters} & 48 (in the case with 3 qubits) or 64 (in the case with 4 qubits) & 64 \\
            \hline
        \end{tabular}
    \end{adjustbox}
    \caption{QNNs properties}
    \label{tab:qnns}
\end{table}

\section{Our Approach} \label{sec:approach}

In this section, we shall present the offers and interaction with a few hardware providers in cloud, evaluated the costs of running our QNNs, and, for those who turned out to be usable, the outcome of running a QNN snippet.

The hardware providers we approached are: IBM Quantum \cite{ibm_quantum}, Amazon Braket \cite{amazon_braket}, Microsoft Azure \cite{azure_quantum}, IQM Quantum Computers \cite{iqm}, Munich Quantum Valley \cite{munich_quantum_valley} and  Pasqal \cite{pasqal}.

All we present in this paper refers to our sole experience in 2024. Table \ref{tab:summary} summarizes our findings in two groups of results: Successful executions, where fidelity and accuracy metrics are available, and Unsuccessful executions, where estimated costs for the sQNN and access issues are exposed.

\begin{table}[htbp]
    \centering
    \renewcommand{\arraystretch}{1.2}
    \begin{tabularx}{\textwidth}{|
        >{\centering\arraybackslash}m{0.20\textwidth}|
        >{\centering\arraybackslash}m{0.20\textwidth}|
        >{\centering\arraybackslash}m{0.11\textwidth}|
        >{\centering\arraybackslash}m{0.17\textwidth}|
        >{\centering\arraybackslash}m{0.18\textwidth}|
    }
        \hline
        \multicolumn{5}{|c|}{\textbf{Successful executions}} \\
        \hline
        \textbf{Hardware used / investigated} &
        \textbf{Circuit type} &
        \textbf{Fidelity / Accuracy} &
        \textbf{Estimated cost for sQNN} &
        \textbf{Response time / Access issues} \\
        \hline
        {\fontfamily{qcr}\selectfont ibm\_kyoto} & \nth{1} order Pauli expansion circuit & 0.995 & Free & None \\
        \hline
        {\fontfamily{qcr}\selectfont ibm\_kyoto} & \nth{1} order Pauli expansion circuit + Pauli Two-Design circuit & 18.701\% & Free & None \\
        \hline
        {\fontfamily{qcr}\selectfont IQM Quantum Computers\_Garnet} & \nth{1} order Pauli expansion circuit & 0.991 & Free & Qiskit version conflicts \\
        \hline
        \multicolumn{5}{|c|}{\textbf{Unsuccessful executions}} \\
        \hline
        \textbf{Hardware used / investigated} &
        \textbf{Circuit type} &
        \textbf{Fidelity / Accuracy} &
        \textbf{Estimated cost for sQNN} &
        \textbf{Response time / Access issues} \\
        \hline
        IBM Quantum QPUs \cite{IBMcomputerslists} & \nth{1} order Pauli expansion circuit + Pauli Two-Design circuit & -- & USD 2000 & No response received \\
        \hline
        IonQ Harmony, IonQ Aria, IQM Garnet, OQC Lucy, QuEra Aquila and Rigetti Aspen-M provided by Amazon Braket & \nth{1} order Pauli expansion circuit + Pauli Two-Design circuit & -- & USD ~58000 & 3 months \\
        \hline
        IonQ Harmony, IonQ Aria and Quantinuum QPUs provided by Microsoft Azure & \nth{1} order Pauli expansion circuit + Pauli Two-Design circuit & -- & Free & Unavailable QPUs \\
        \hline
        IonQ Harmony, IonQ Aria provided by Microsoft Azure & \nth{1} order Pauli expansion circuit + Pauli Two-Design circuit & -- & Starting from USD 5,000,000 or USD 25,000 per month & -- \\
        \hline
        Quantinuum QPUs provided by Microsoft Azure & \nth{1} order Pauli expansion circuit + Pauli Two-Design circuit & -- & Starting from USD 125,000 per month & -- \\
        \hline
        Pasqal QPUs  & \nth{1} order Pauli expansion circuit + Pauli Two-Design circuit & -- & Unknown & No response received \\
        \hline
        Munich Quantum Valley QPUs  & \nth{1} order Pauli expansion circuit + Pauli Two-Design circuit & -- & Unknown & No response received \\
        \hline
    \end{tabularx}
    \caption{The summary of our experience investigating the prospects of running the sQNN on quantum hardware. The first section (Successful executions) presents the results we obtained running a segment of the sQNN on some quantum computers, such as fidelity and accuracy. The second section (Unsuccessful executions) exposes the impediments we met when we approached the quantum hardware providers, such as very high costs and access issues.}
    \label{tab:summary}
\end{table}

\subsection{IBM Quantum} \label{sec:IBMQuantum}

In 2024, IBM Quantum \cite{ibm_quantum} had the \textit{Open Plan} featuring 10 minutes of running time per month and access to 4 quantum computers: {\fontfamily{qcr} \selectfont ibm\_sherbrooke, ibm\_brisbane, ibm\_osaka} and {\fontfamily{qcr} \selectfont ibm\_kyoto}. Pricing plans were also listed. The \textit{Pay-as-you-go} plan enables access to more quantum computers (such as {\fontfamily{qcr} \selectfont ibm\_kyiv, ibm\_strasbourg,} etc. \cite{IBMcomputerslists} and IBM Quantum technical support for the price of USD 1.60 per second of hardware usage. There are two other subscription based plans, with variable costs: the \textit{Premium Plan} and the \textit{Dedicated Service}. These plans provide additional benefits, including IBM Quantum Network membership, access to the Quantum Accelerator offering, and the option for system deployment at the client’s site. In order to get an offer, one must fill in an application form. We submitted three application 
forms, which, for reasons unknown to us, did not receive a response to the time of the publication of this work.

\begin{comment}
\begin{figure}
    \centering
    \includegraphics[width=\linewidth]{IBM_Open_Plan.png}
    \caption{Open Plan of IBM Quantum}
    \label{fig:IBM_Open_Plan}
\end{figure}

\begin{figure}
    \centering
    \includegraphics[width=\linewidth]{IBM_pricing_plans.png}
    \caption{Pricing plans offed by IBM Quantum}
    \label{fig:IBM_pricing_plans}
\end{figure}

\begin{figure}
    \centering
    \includegraphics[width=\linewidth]{IBM_application_form.png}
    \caption{The form one must fill in order to apply for a pricing plan at IBM in 2024. We sent three applications and never received an answer.}
    \label{fig:IBM_application_form}
\end{figure}
\end{comment}

\begin{figure}
    \centering
    \includegraphics[width=\linewidth]{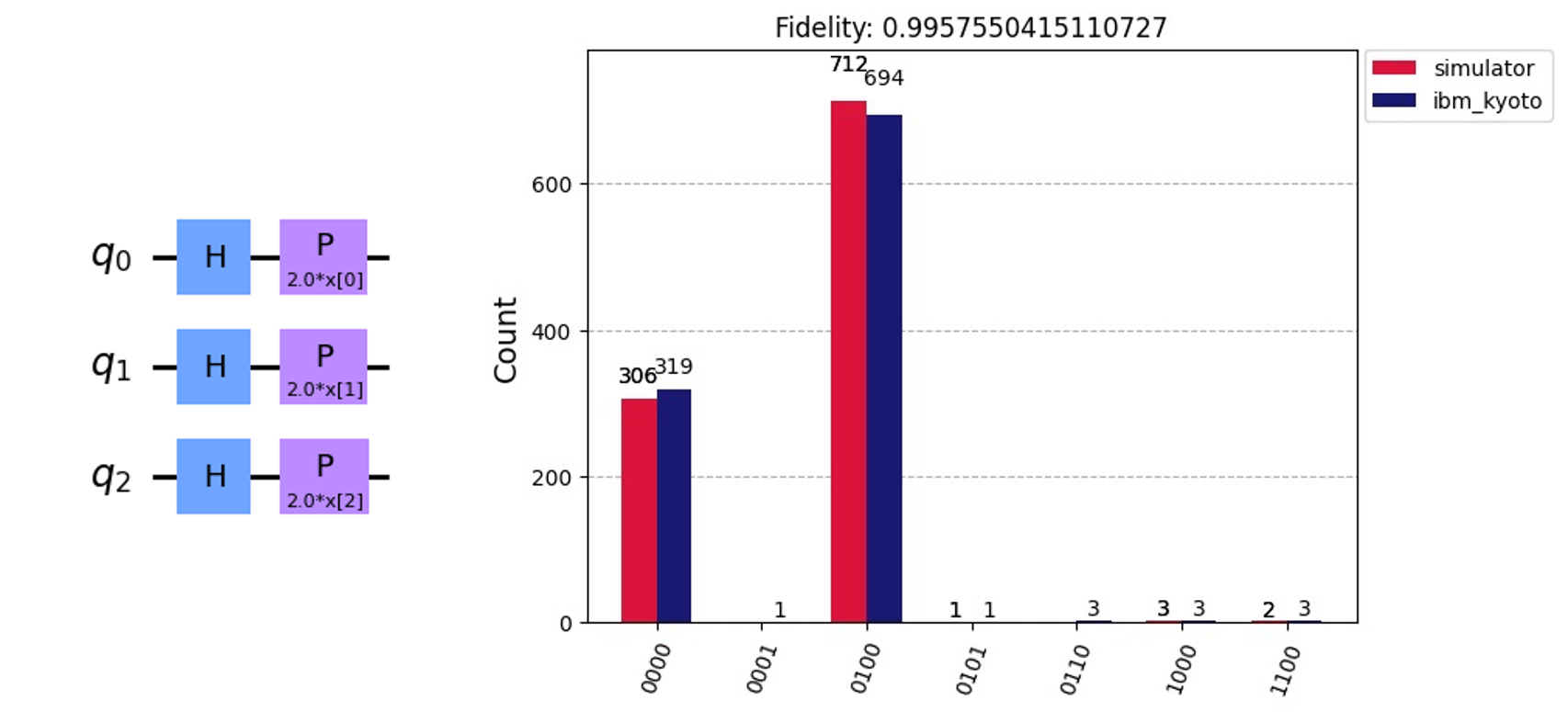}
    \caption{The quantum circuit we ran on {\fontfamily{qcr}\selectfont ibm\_kyoto} (left) and the obtained fidelity (right).}
    \label{fig:IBM_run_1}
\end{figure}

Using the \textit{Open Plan}, we ran the feature map of the sQNN (Figure\ref{fig:IBM_run_1}) on {\fontfamily{qcr} \selectfont ibm\_kyoto}, with optimization level 3 and 1024 shots. Transpilation was performed implicitly by Qiskit, and the fully transpiled circuits are not exposed to the user. We evaluated the fidelity of the outcome for one random sQNN input entry using the square of the Bhattacharyya coefficient \cite{fidelity1}, \cite{fidelity2} computed for the probability density simulated on a local classical computer and the probability density obtained with {\fontfamily{qcr} \selectfont ibm\_kyoto}. The resulting fidelity was 0.995, indicating excellent agreement between the simulated and experimental probability density distributions (Figure\ref{fig:IBM_run_1}).

\begin{figure}
    \centering
    \includegraphics[width=\linewidth]{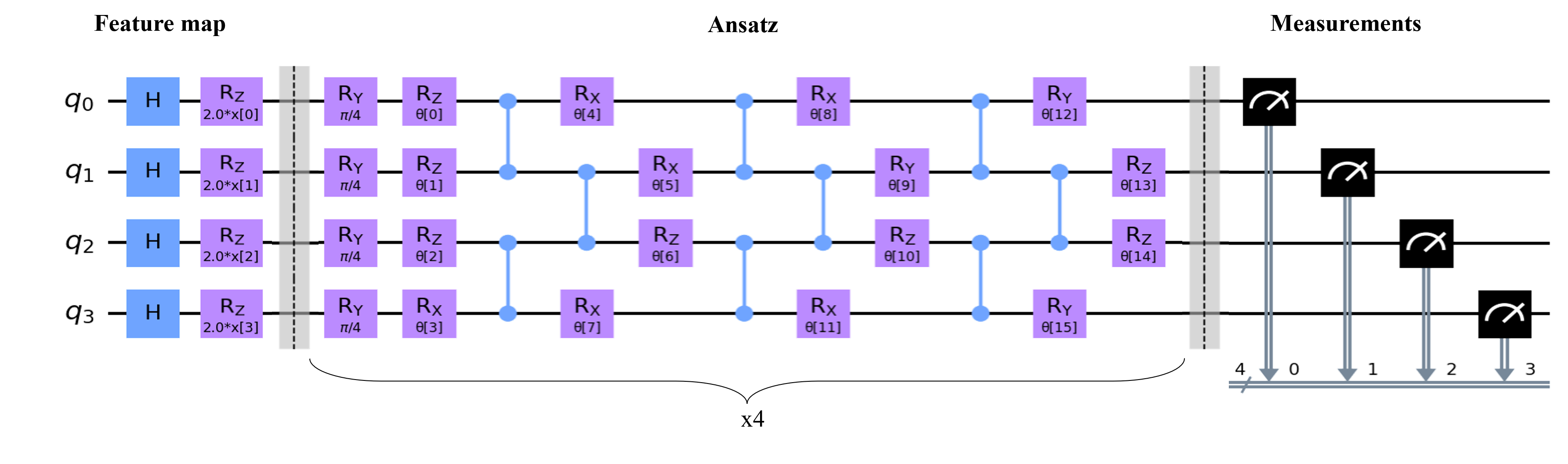}
    \caption{The quantum circuit (part of the sQNN) that was ran on {\fontfamily{qcr}\selectfont ibm\_kyoto} for making predictions on 1802 samples of data. The classification accuracy was 18.701\%, while on the simulator it was 100\%.}
    \label{fig:IBM_sQNN}
\end{figure}

\begin{figure}[htbp]
    \centering
    \begin{subfigure}[t]{0.45\textwidth}
        \centering
        \includegraphics[width=\textwidth]{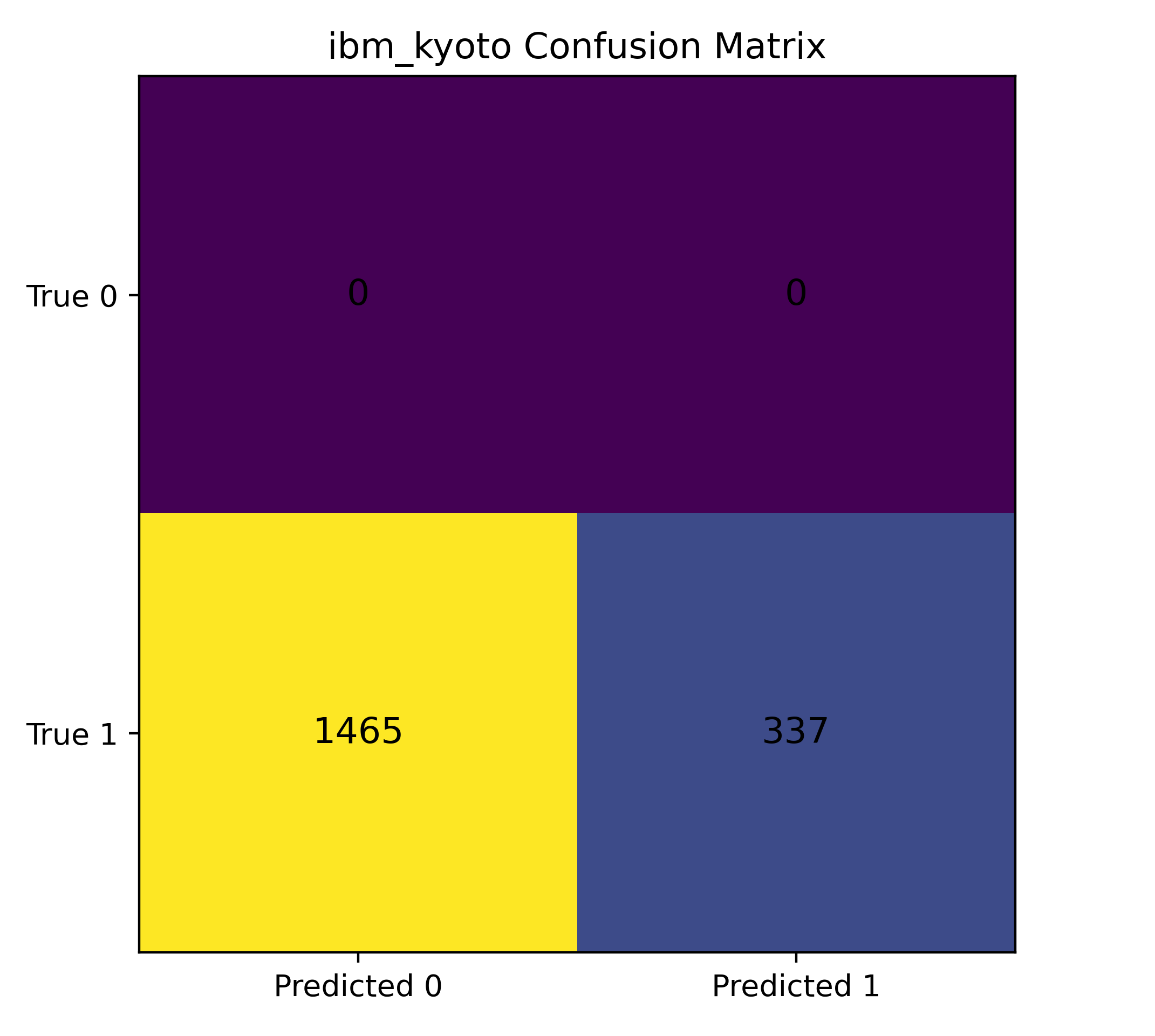}
        \caption{}
        \label{fig:conf_matrix_ibm_kyoto}
    \end{subfigure}
    \hspace{0.01\textwidth}
    \begin{subfigure}[t]{0.45\textwidth}
        \centering
        \includegraphics[width=\textwidth]{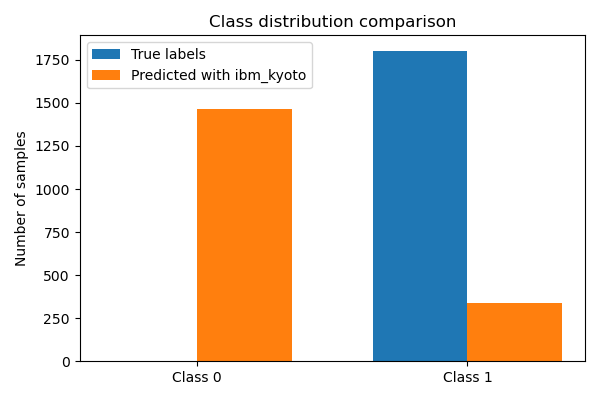}
        \caption{}
        \label{fig:class_dist_comparison_ibm_kyoto}
    \end{subfigure}

    \vspace{0.5cm}

    \begin{subfigure}[t]{0.30\textwidth}
        \centering
        \includegraphics[width=\textwidth]{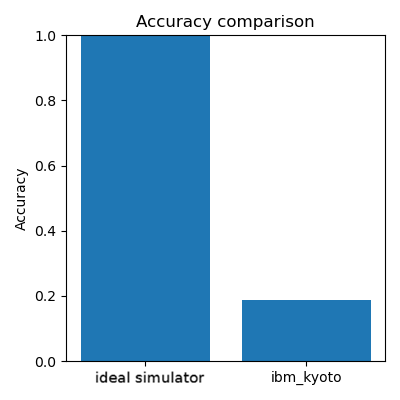}
        \caption{}
        \label{fig:acc_comparison_ibm_kyoto}
    \end{subfigure}
    \hspace{0.15\textwidth}
    \begin{subfigure}[t]{0.45\textwidth}
        \centering
        \includegraphics[width=\textwidth]{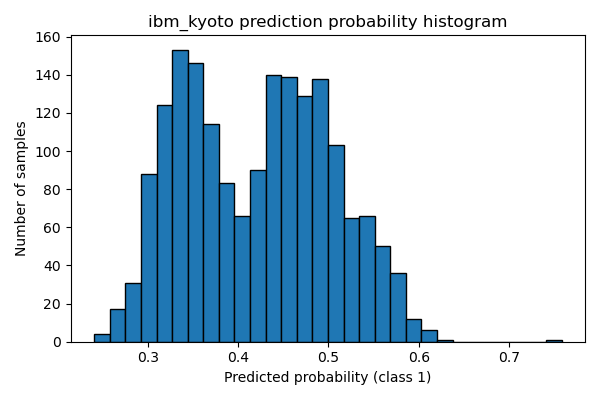}
        \caption{}
        \label{fig:pred_prob_hist_ibm_kyoto}
    \end{subfigure}

    \caption{Results obtained when executing the sQNN for prediction making with {\fontfamily{qcr} \selectfont ibm\_kyoto}, showing that the learned decision boundary is not preserved and the hardware induces a strong bias towards class 0. Panel (a): Confusion matrix for predictions obtained with {\fontfamily{qcr} \selectfont ibm\_kyoto}. Panel (b): Class distribution comparison between the true labels and labels obtained with {\fontfamily{qcr} \selectfont ibm\_kyoto}. Panel (c): Accuracy comparison between ideal accuracy and the accuracy obtained with {\fontfamily{qcr} \selectfont ibm\_kyoto}. Panel (d): Prediction probability histogram obtained with {\fontfamily{qcr} \selectfont ibm\_kyoto}.}
    \label{fig:ibm_kyoto_results}
\end{figure}

Next, we decided to make predictions with the sQNN on 1802 samples of data, previously trained on our laptops, with {\fontfamily{qcr} \selectfont ibm\_kyoto}. The dataset was split in several chunks in order to be possible to execute the circuit (Fig.\ref{fig:IBM_sQNN}) using only the \textit{Open Plan}, then the outcomes were concatenated. The classification accuracy was 18.701\%, while on the ideal simulator it was 100\% (Figure \ref{fig:acc_comparison_ibm_kyoto}). The {\fontfamily{qcr} \selectfont ibm\_kyoto} quantum computer introduces a bias towards class 0, as it can be seen in Figure \ref{fig:class_dist_comparison_ibm_kyoto} and in Figure \ref{fig:pred_prob_hist_ibm_kyoto}. While the true labels are all 1, the hardware predictions are strongly skewed towards  class 0. This indicates that the learned decision boundary is not preserved when executing the sQNN with {\fontfamily{qcr} \selectfont ibm\_kyoto}, leading to a lot of false negative predictions (Figure \ref{fig:conf_matrix_ibm_kyoto}).

\subsection{Amazon Braket}

Amazon Braket \cite{amazon_braket} offers cloud access to some quantum computers, owned by different companies, such as IonQ Harmony, IonQ Aria \cite{ionq}, IQM \cite{iqm}, Rigetti, QuEra and OQC. There is a free service which consists of 1h/month free access to simulators. For quantum computers, prices are per-task (circuit) and per-shot. For any quantum computer, the per-task price is USD 0.3, while the per-shot price varies between USD 0.00035 and USD 0.03. As the sQNN requires $\approx$160 training epochs and using 2500 shots per task, the estimated cost for running once is approximately USD 58000. In order to access the quantum computers, one must create a reservation, which is manually managed and an answer is guaranteed in 2-3 business days. We received an answer after circa 3 months.

\begin{comment}
\begin{figure}
    \centering
    \includegraphics[width=\linewidth]{amazon_braket_quantum_computers.png}
    \caption{Quantum processors accessible in cloud via Amazon Braket and costs}
    \label{fig:amazon_braket_quantum_computers}
\end{figure}

\begin{figure}
    \centering
    \includegraphics[width=\linewidth]{amazon_braket_response.png}
    \caption{Amazon Braket message after creating a reservation for using one of the quantum processors they provide cloud access to.}
    \label{fig:amazon_braket_response}
\end{figure}
\end{comment}

\subsection{Microsoft Azure}

Microsoft Azure \cite{azure_quantum} provides cloud access to several quantum computers owned by companies such as IonQ Harmony, IonQ Aria \cite{ionq}, Quantinuum \cite{azure_quantum}, Pasqal \cite{pasqal}, etc. The free plan includes 500 or 200 credits for free, depending on the account, which cover the execution of one approximately 100 depth circuit. Fees are per gate and shot, ranging from USD 0.00003 to USD 0.000975 for IonQ quantum computers. The estimated costs for one run the sQNN, which requires 160 training epochs, and considering 2,500 shots, start from USD 5,000,000. Also, there are monthly subscriptions available for IonQ and Quantinuum quantum computers at the price of USD 2500/month or USD 185000/month, plus additional Azure infrastructure costs.

We tried multiple times to use the free credits and access IonQ and Quantinuum processors, but the backends were never available.

\begin{comment}
\begin{figure}
    \centering
    \includegraphics[width=\linewidth]{microsoft_azure_costs.png}
    \caption{Pricing for the cloud access to quantum computers provided by Microsoft Azure}
    \label{fig:microsoft_azure_costs}
\end{figure}

\begin{figure}
    \centering
    \includegraphics[width=\linewidth]{ionq_never_available.png}
    \caption{The type of error we get when we try to access an IonQ quantum processor using the Microsoft Azure platform. The same error appeared when trying to use the Quantinuum quantum processors.}
    \label{fig:ionq_never_available}
\end{figure}

\begin{figure}
    \centering
    \includegraphics[width=\linewidth]{pasqal_form.png}
    \caption{The form requested by Pasqal in order to start using their quantum processors. We filled and sent it two times, but we never received an answer.}
    \label{fig:pasqal_form}
\end{figure}
\end{comment}

Pasqal quantum computers could be accessed through Microsoft Azure, but in reality a separate, direct form must be filled. We filled it and did not receive an answer from Pasqal to the publication of this paper.

\subsection{IQM Quantum Computers}

IQM Resonance \cite{iqm} is the quantum computer cloud access provided by IQM Quantum Computers \cite{iqm}. In order to access it, one must fill a form on their site. They offer a free hour of quantum computer usage. We managed to get access to {\fontfamily{qcr}\selectfont IQM Quantum Computers\_Garnet} quantum computer after the second form we sent. One hour would have been enough to run the sQNN, both the training and the prediction parts, but we encountered a Qiskit version conflict error. The {\fontfamily{qcr}\selectfont IQM Quantum Computers\_Garnet} compiler required Qiskit version is older than any version the sQNN is able to run with, so we were able to run only the feature map, with optimization level 3 and 1024 shots. The fully transpiled circuits are not exposed to the user, since transpilation was performed implicitly by Qiskit,. The fidelity, evaluated for one randum sQNN input entry as the square of the Bhattacharyya coefficient, is 0.991.

\begin{comment}

\begin{figure}
    \centering
    \includegraphics[width=\linewidth]{iqm_form.png}
    \caption{IQM Quantum Computers form for quantum computer cloud access.}
    \label{fig:IQM Quantum Computers_form}
\end{figure}

\begin{figure}
    \centering
    \includegraphics[width=\linewidth]{iqm_error.png}
    \caption{The Qiskit version conflict error we get when trying to run one of our QNNs on IQM Quantum Computers's {\fontfamily{qcr}\selectfont IQM Quantum Computers\_Garnet}, which stopped us from using it.}
    \label{fig:IQM Quantum Computers_error}
\end{figure}

\begin{figure}
    \centering
    \includegraphics[width=\linewidth]{iqm_results.png}
    \caption{The circuit we were able to run on the IQM Quantum Computers quantum computer (left) and the fidelity (right).}
    \label{fig:IQM Quantum Computers_results}
\end{figure}

\end{comment}

\subsection{Munich Quantum Valley}

In order to start working with Munich Quantum Valley \cite{munich_quantum_valley} quantum computers, one needed to sign up to their portal and request access. We filled in the form, and as of 15 November 2025 no answer has been received.

\begin{comment}
    \begin{figure}
        \centering
        \includegraphics[width=\linewidth]{munich_quantum_valley_form.png}
        \caption{The form for requesting access to the Munich Quantum Valley quantum computers}
    \label{fig:munich_quantum_valley_form}
\end{figure}
\end{comment}

\section{Qiskit Issues} \label{sec:qiskit}

We used Qiskit to program our QNNs. This library is easy to use and widely accepted by hardware providers. However, we encountered a lot of difficulties when actually trying to run a circuit written in Qiskit. The main issues were the following:

\begin{itemize}
    \item Version incompatibility between {\fontfamily{qcr}\selectfont
qiskit} and {\fontfamily{qcr}\selectfont qiskit-ibm-runtime}
    \item Deprecations, for instance: functions removed before documentation for the new replacements were complete
    \item Provider-specific constraints: unavailable Qiskit versions required by IQM compilers and for IBM runtimes
\end{itemize}

In 2024, there were always conflicts or incompatibilities between the {\fontfamily{qcr}\selectfont
qiskit} and {\fontfamily{qcr}\selectfont qiskit-ibm-runtime} packages, as frequent deprecations occurred and new versions emerged. In most cases, the {\fontfamily{qcr}\selectfont qiskit} last version was deprecated and a new version was launched; this new version lacked proper documentation and a compatible {\fontfamily{qcr}\selectfont qiskit-ibm-runtime} package. Also, the old version was not compatible anymore with {\fontfamily{qcr}\selectfont qiskit-ibm-runtime}. This situation led to very slow debugging in order to be able to run code locally and the necessity of allowing about a week in order to get a functional {\fontfamily{qcr}\selectfont qiskit-ibm-runtime} package and be able to access a quantum computed in cloud. Occasionally, there were frustrating moments of blind debugging because Qiskit kept deprecating and launching new versions very fast, while {\fontfamily{qcr}\selectfont qiskit-ibm-runtime} remained unusable. Once, when we wanted to run the second time a quantum circuit with IQM Quantum Computers, the Qiskit version that allows running on quantum computers was not supported by IQM Quantum Computers compilers.

\section{Discussion} 

The pay-as-you-go plan is available for most hardware providers and can be used as the fees as billed directly to the user and there is no need for an application. Shallow quantum circuits can run with high fidelity ($>90\%$) and reasonable costs at hardware providers like IBM Quantum, Microsoft Azure \cite{azure_quantum}, Amazon Braket \cite{amazon_braket}, IQM Quantum Computers \cite{iqm}. For deeper circuits, as the {\fontfamily{qcr} \selectfont ibm\_kyoto} results presented in Section \ref{sec:IBMQuantum}, the fidelity drops, making it pointless to run one sQNN epoch.

Regarding providers such as IonQ, Microsoft Azure, Amazon Braket and Quantinuum, the very high fees make it really difficult to run the whole sQNN, let alone the lQNN.

For other providers, like Munich Quantum Valley \cite{munich_quantum_valley} and Pasqal \cite{pasqal}, it was not possible to evaluate any of their fees or hardware capabilities because they never replied to application forms to their plans.

A major issue that slows the process of using a quantum computer in cloud is the Qiskit versions conflicts. There are frequent conflicts between the {\fontfamily{qcr}\selectfont qiskit} and {\fontfamily{qcr}\selectfont qiskit-ibm-runtime} packages, conflicts between the Qiskit version requested by the provider and the Qiskit version that accepts real quantum computers as backends. To this adds the deprecation of past Qiskit version and the lack of documentation for the new versions. Plus, the debugging process is slower than the new Qiskit versions releases.

Thus, these impediments can be separated in two categories: hardware limitations and system limitations. The hardware limitations refer to all hardware flaws, such as readout errors and decoherence, which lead to unreliable results as seen in Section \ref{sec:IBMQuantum}, while the system limitations include pricing models, access control, and software stack (Qiskit/runtime compatibilities). Following this discussion, several specific recommendations can be made. For QML practitioners, circuits should be kept shallow; toy-scale demonstrations should be prioritized on hardware, while full model training should rely on simulators. For hardware providers, improvements are needed in the alignment of SDK and runtime versions, in the transparency and speed of access processes, and in the development of pricing models tailored to QML workloads, which typically involve many short circuits and a large number of shots. Finally, for LISA data analysis, the near-term role of quantum hardware should be regarded as exploratory rather than operational.

\section{Conclusion}

The current landscape of cloud-based quantum computing for QML workloads remains highly constrained by both hardware limitations and system-level limitations. While shallow circuit executions turn out with high fidelity, deeper circuits and full-scale training are impractical due to rapid fidelity degradation and prohibitive costs. System challenges, such as SDK/runtime incompatibilities, Qiskit version conflicts, and opaque access processes, further hinder usability. Consequently, near-term quantum hardware should be viewed only as a tool for proof-of-concept or exploratory experiments rather than operational deployment. Progress will depend on coordinated improvements in hardware reliability, pricing tailored to QML workloads, and streamlined software ecosystems. Until these issues are addressed, simulators remain the primary support for meaningful QML development.

\section*{Aknowledgements}

We acknowledge the use of IBM Quantum services for this work. The views expressed are those of the authors, and do not reflect the official policy or position of IBM or the IBM Quantum team.

\bibliographystyle{alpha}
\bibliography{bib}

\end{document}